\documentclass[runningheads]{llncs}
\usepackage{graphicx}

\usepackage[hang,flushmargin]{footmisc}
\usepackage[misc,geometry]{ifsym}
\usepackage{cite} 
\usepackage[T1]{fontenc}
\usepackage{amsmath}
\usepackage{mathrsfs}
\usepackage{graphicx}  
\usepackage{epstopdf}
\usepackage{bbm}
\usepackage{float}
\usepackage{multirow}
\usepackage{booktabs}
\usepackage{mathrsfs}
\usepackage{amssymb}
\usepackage[colorlinks,linkcolor=red]{hyperref}
\usepackage{siunitx}
\makeatletter

\newcommand{\Rmnum}[1]{\expandafter\@slowromancap\romannumeral #1@}
\makeatother
\usepackage{booktabs}
\usepackage{amsfonts,amssymb}
\usepackage{amsmath,bm}
\usepackage{array}
\newcommand{\PreserveBackslash}[1]{\let\temp=\\#1\let\\=\temp}
\newcolumntype{C}[1]{>{\PreserveBackslash\centering}p{#1}}
\newcolumntype{R}[1]{>{\PreserveBackslash\raggedleft}p{#1}}
\newcolumntype{L}[1]{>{\PreserveBackslash\raggedright}p{#1}}
\usepackage{algorithmicx,algorithm}
\makeatletter

\newcommand{\para}[1]{\vspace{.05in}\noindent\textbf{#1}}

\def\etal{{\em et al.}}

\usepackage{xcolor}

\begin{document}

\title{Make-A-Volume: Leveraging Latent \\Diffusion Models for Cross-Modality 3D Brain MRI Synthesis}

\titlerunning{Make-A-Volume}

\author{
Lingting Zhu\inst{1}\and
Zeyue Xue\inst{1} \and
Zhenchao Jin\inst{1} \and
Xian Liu\inst{2} \and \\
Jingzhen He\inst{3}\textsuperscript{(\Letter)} \and
Ziwei Liu\inst{4} \and
Lequan Yu\inst{1}\textsuperscript{(\Letter)}
}
%index{Zhu, Lingting}
%index{Xue, Zeyue}
%index{Jin, Zhenchao}
%index{Liu, Xian}
%index{He, Jingzhen}
%index{Liu, Ziwei}
%index{Yu, Lequan}

\authorrunning{L. Zhu et al.}
\institute{
    The University of Hong Kong, Hong Kong SAR, China \\ \email{ltzhu99@connect.hku.hk lqyu@hku.hk} \and
    The Chinese University of Hong Kong, Hong Kong SAR, China \and
    Qilu Hospital of Shandong University, Jinan, China \\\email{hjzhhjzh@163.com} \and
    S-Lab, Nanyang Technological University, Singapore
}
\maketitle

\begin{abstract}

Cross-modality medical image synthesis is a critical topic and has the potential to facilitate numerous applications in the medical imaging field.
Despite recent successes in deep-learning-based generative models, most current medical image synthesis methods rely on generative adversarial networks and suffer from notorious mode collapse and unstable training. 
Moreover, the 2D backbone-driven approaches would easily result in volumetric inconsistency, while 3D backbones are challenging and impractical due to the tremendous memory cost and training difficulty.
In this paper, we introduce a new paradigm for volumetric medical data synthesis by leveraging 2D backbones and present a diffusion-based framework, \textbf{Make-A-Volume}, for cross-modality 3D medical image synthesis.
To learn the cross-modality slice-wise mapping, we employ a latent diffusion model and learn a low-dimensional latent space, resulting in high computational efficiency.
To enable the 3D image synthesis and mitigate volumetric inconsistency, we further insert a series of volumetric layers in the 2D slice-mapping model and fine-tune them with paired 3D data.
This paradigm extends the 2D image diffusion model to a volumetric version with a slightly increasing number of parameters and computation, offering a principled solution for generic cross-modality 3D medical image synthesis.
We showcase the effectiveness of our Make-A-Volume framework on an in-house SWI-MRA brain MRI dataset and a public T1-T2 brain MRI dataset. 
Experimental results demonstrate that our framework achieves superior synthesis results with volumetric consistency.


\end{abstract}
\section{Introduction}
\label{sec:intro}
Medical images are essential in diagnosing and monitoring various diseases and patient conditions.
Different imaging modalities, such as computed tomography (CT) and magnetic resonance imaging (MRI), and different parametric images, such as T1 and T2 MRI, have been developed to provide clinicians with a comprehensive understanding of the patients from multiple perspectives~\cite{filippou2018recent}. 
However, in clinical practice, it is commonly difficult to obtain a complete set of multiple modality images for diagnosis and treatment due to various reasons, such as modality corruption, incorrect machine settings, allergies to specific contrast agents, and limited available time~\cite{dar2019image, hu2022autogan}.
Therefore, cross-modality medical image synthesis is useful by allowing clinicians to acquire different characteristics across modalities and facilitating real-world applications in radiology and radiation oncology~\cite{yu2019ea,wang2021review}.

With the rise of deep learning, numerous studies have emerged and are dedicated to medical image synthesis~\cite{nie2018medical, filippou2018recent, dalmaz2022resvit}. 
Notably, generative adversarial networks (GANs)~\cite{goodfellow2020generative} based approaches have garnered significant attention in this area due to their success in image generation and image-to-image translation~\cite{isola2017image, zhu2017unpaired}. 
Moreover, GANs are also closely related to cross-modality medical image synthesis~\cite{ben2019cross, yu2019ea, hu2022autogan}.
However, despite their efficacy, GANs are susceptible to mode collapse and unstable training, which can negatively impact the performance of the model and decrease the reliability in practice~\cite{li2018implicit, bau2019seeing}. 
Recently, the advent of denoising diffusion probabilistic models (DDPMs)~\cite{sohl2015deep, ho2020denoising} has introduced a new scheme for high-quality generation, offering desirable features such as better distribution coverage and more stable training when compared to GAN-based counterparts. 
Benefiting from the better performance~\cite{dhariwal2021diffusion}, diffusion-based models may be deemed much more reliable and dominant and recently researchers have made the first attempts to employ diffusion models for medical image synthesis~\cite{pinaya2022brain, kim2022diffusion, khader2022medical, kazerouni2022diffusion}. 

Different from natural images, most medical images are volumetric. 
Previous studies employ 2D networks as backbones to synthesize slices of medical volumetric data due to their ease of training~\cite{nie2018medical, yu2019ea} and then stack 2D results for 3D synthesis. 
However, this fashion induces volumetric inconsistency, particularly along the z-axis when following the standard way of placing the coordinate system. 
Although training 3D models may avoid this issue, it is challenging and impractical due to the massive amount of volumetric data required, and the higher dimension of the data would result in costly memory requirements~\cite{uzunova2020memory, chung2022solving, lee2023improving}. 
To sum up, balancing the trade-off between training and volumetric consistency remains an open question that requires further investigation.

In this paper, we propose \textbf{Make-A-Volume}, a diffusion-based pipeline for cross-modality 3D brain MRI synthesis. 
Inspired by recent works that factorize video generation into multiple stages~\cite{singer2022make, wu2022tune}, 
we introduce a new paradigm for volumetric medical data synthesis by leveraging 2D backbones to simultaneously facilitate high-fidelity cross-modality synthesis and mitigate volumetric inconsistency for medical data. 
Specifically, we employ a latent diffusion model (LDM)~\cite{rombach2022high} to function as a slice-wise mapping that learns cross-modality translation in an image-to-image manner.
Benefiting from the low-dimensional latent space of LDMs, the high memory requirements for training are mitigated.
To enable the 3D image synthesis and enhance volumetric smoothness among medical slices, we further insert and fine-tune a series of volumetric layers to upgrade the slice-wise model to a volume-wise model. 
In summary, our contributions are three-fold: 
(1) We introduce a generic paradigm for 3D image synthesis with 2D backbones, which can mitigate volumetric inconsistency and training difficulty related to 3D backbones.
(2) We propose an efficient latent diffusion-based framework for high-fidelity cross-modality 3D medical image synthesis.
(3) We collected a large-scale high-quality dataset of paired susceptibility weighted imaging (SWI) and magnetic resonance angiography (MRA) brain images. 
Experiments on these in-house and public T1-T2 brain MRI datasets show the volumetric consistency and superior quantitative result of our framework.
\section{Method}

\begin{figure}[t]
\includegraphics[width=\textwidth]{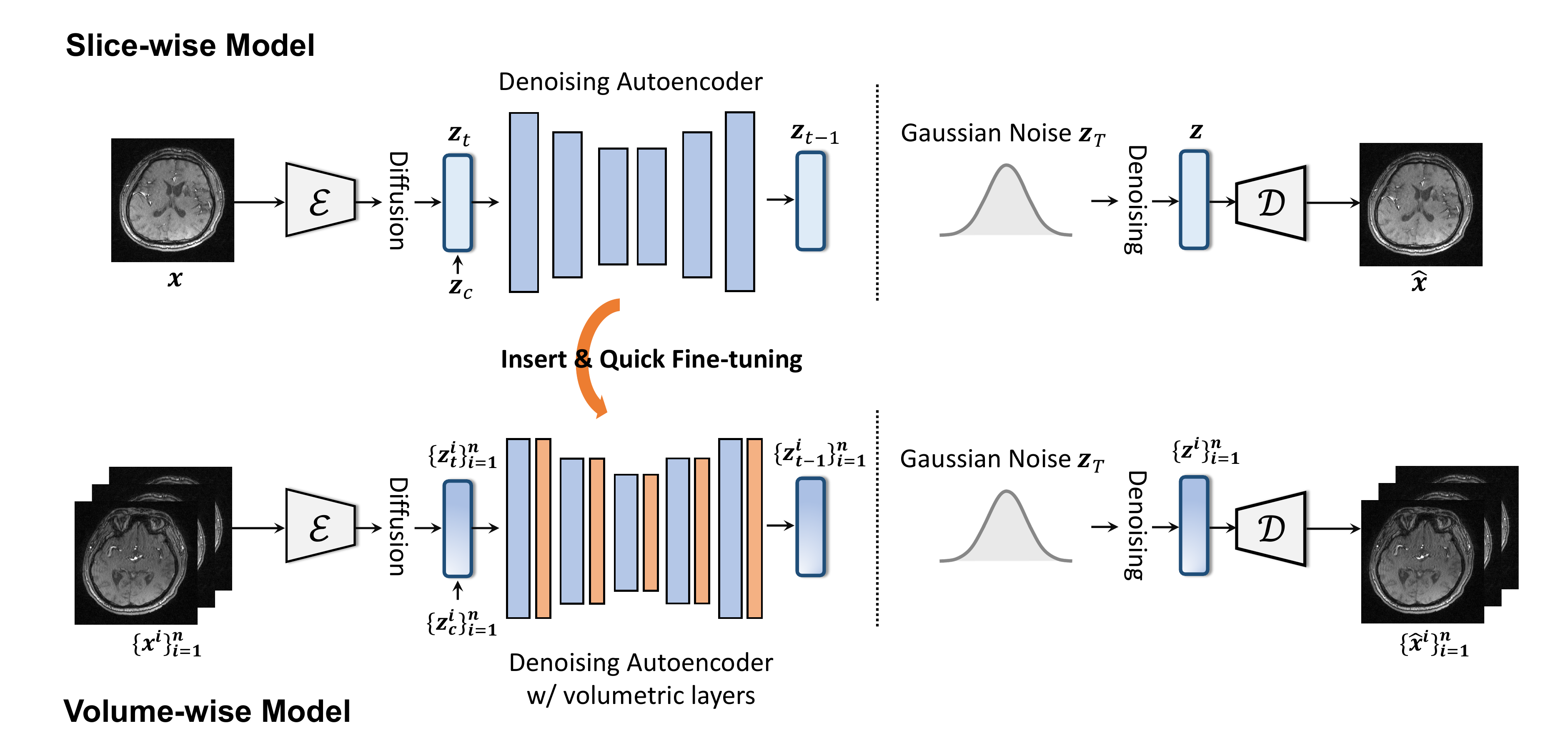}
\caption{\textbf{Overview of our proposed two-stage Make-A-Volume framework.} 
A latent diffusion model is used to predict the noises added to the image and synthesize independent slices from Gaussian noises. 
We insert volumetric layers and quickly fine-tune the model, which extends the slice-wise model to be a volume-wise model and enables synthesizing volumetric data from Gaussian noises.} 
\label{overview}
\vspace{-0.3cm}
\end{figure}

\vspace{-0.2cm}
\subsection{Preliminaries of DDPMs}
In the diffusion process, DDPMs produce a series of noisy inputs ${x_0, x_1, ..., x_T}$, via sequentially adding Gaussian noises to the sample over a predefined number of timesteps $T$. Formally, given clean data samples which follow the real distribution $x_0 \sim q(x)$, the diffusion process can be written down with variances $\beta_1, ..., \beta_T$ as 
\begin{align} \label{eq:forward}
q(x_t | x_{t-1}) &= \mathcal{N}(x_t; \sqrt{1 - \beta_t}x_{t-1}, \beta_t \mathbf{I}).
\end{align} 
Employing the property of DDPMs, the corrupted data $x_t$ can be sampled easily from $x_0$ in a closed form:
\begin{align} \label{eq:sample}
q(x_t | x_0) = \mathcal{N}(x_t; \sqrt{\bar{\alpha}_t}x_0, (1 - \bar{\alpha}_t) \mathbf{I}); \ \ x_t = \sqrt{\bar{\alpha}_t}x_0 + \sqrt{1 - \bar{\alpha}_t} \epsilon,
\end{align} 
where $\alpha_t = 1 - \beta_t$, $\bar{\alpha}_t = \prod_{s=1}^t \alpha_s$, and $\epsilon \sim \mathcal{N}(0, 1)$ is the added noise.

In the reverse process, the model learns a Markov chain process to convert the Gaussian distribution into the real data distribution by predicting the parameterized Gaussian transition $p(x_{t-1} | x_t)$ with the learned model $\theta$:
\begin{align} \label{eq:reverse}
p_{\theta}(x_{t-1}|x_t) &= \mathcal{N}(x_{t-1}; \mu_{\theta}(x_t, t), \sigma^2_t\mathbf{I}).
\end{align} 

In the model training, the model tries to predict the added noise $\epsilon$ with the simple mean squared error (MSE) loss: 
\begin{align}
    L (\theta) = \mathbb{E}_{x_0 \sim q(x), \epsilon \sim \mathcal{N}(0, 1), t}\left[\left\| \epsilon - \epsilon_\theta(\sqrt{\bar\alpha_t} x_0 + \sqrt{1-\bar\alpha_t}\epsilon, t) \right\|^2\right].
\label{eq:losssimple}
\end{align}

\subsection{Slice-wise Latent Diffusion Model} 
To improve the computational efficiency of DDPMs that learn data in pixel space,  Rombach~\etal~\cite{rombach2022high} proposes training an autoencoder with a KL penalty or a vector quantization layer~\cite{kingma2013auto, van2017neural}, and introduces the diffusion model to learn the latent distribution. 
Given calibrated source modality image $x_c$ and target modality image $x$, we leverage a slice-wise latent diffusion model to learn the cross-modality translation. 
With the pretrained encoder $\mathcal{E}$, $x_c$ and $x$ are compressed into a spatially lower-dimensional latent space of reduced complexity, generating $z_c$ and $z$. 
The diffusion and denoising processes are then implemented in the latent space and a U-Net~\cite{ronneberger2015u} is trained to predict the noise in the latent space. 
The input consists of the concatenated $z_c$ and $z$ and the network learns the parameterized Gaussian transition $p_{\theta}(z_{t-1}|z_t, z_c) = \mathcal{N}(z_{t-1}; \mu_{\theta}(z_t, t, z_c), \sigma^2_t \mathbf{I})$. 
After learning the latent distribution, the slice-wise model can synthesize target latent $\hat{z}$ from Gaussian noise, given the source latent $z_c$. Finally, the decoder $\mathcal{D}$ restores the slice to the image space via $\hat{x} = \mathcal{D}(\hat{z})$. 

\subsection{From Slice-wise Model to Volume-wise Model}
Fig.~\ref{overview} illustrates an overview of the Make-A-Volume framework. 
The first stage involves a latent diffusion model that learns the cross-modality translation in an image-to-image manner to synthesize independent slices from Gaussian noises. Then, to extend the slice-wise model to be a volume-wise model, we insert volumetric layers and quickly fine-tune the U-Net. As a result, the volume-wise model synthesizes volumetric data without inconsistency from Gaussian noises.

In the slice-wise model, distribution of the latent $z \in \mathbb{R}^{b_s \times c\times h\times w}$ is learned by the U-Net, where $b_s, c, h, w$ are the batch size of slice, channels, height, and width dimensions respectively, and there is where little volume-awareness is introduced to the network. 
Since we target in synthesizing volumetric data and assume each volume consists of $N$ slices, we can factorize the batch size of slices as $b_s = b_v n$, where $B_v$ represents the batch size of volumes. Now, volumetric layers are injected and help the U-Net learn to latent feature $f \in \mathbb{R}^{(b_v\times n) \times c\times h\times w}$ with volumetric consistency. The volumetric layers are basic 1D convolutional layers and the $i-$th volumetric layer $l^i_v$ takes in feature $f$ and outputs $f^{\prime}$ as:

\begin{align} \label{lv}
&f^{\prime} \leftarrow \text{Rearrange}(f, (b_v\times n)\ c\ h\ w \rightarrow (b_v\times h\times w)\ c\ n),\\
&f^{\prime} \leftarrow l^i_v(f^{\prime}),\\
&f^{\prime} \leftarrow \text{rearrange}(f, (b_v\times h\times w)\ c\ n \rightarrow (b_v\times n)\ c\ h\ w).
\end{align} 

Here, the 1D conv layers combined with the pretrained 2D conv layers, serve as pseudo 3D conv layers with little extra memory cost. 
We initialize the volumetric 1D convolution layers as Identity Functions for more stable training and we empirically find tuning is efficient. 
With the volume-aware network, the model learns volume data $\{x^i\}^n_{i=1}$, predicts $\{z^i\}^n_{i=1}$, and reconstruct $\{{\hat{x}}^i\}^n_{i=1}$.
For diffusion model training, in the first stage, we randomly sample timestep $t$ for each slice. However, when tuning the second stage, the U-Net with volumetric layers learns the relationship between different slices in one volume. As a result, fixing $t$ for each volume data is necessary and we encourage the small $t$ values to be sampled more frequently for easy training. In detail, we sample the timestep $t$ with replacement from multinomial distribution, and the pre-normalized weight (used for computing probabilities after normalization) for timestep $t$ equals $2T-t$, where $T$ is the total number of timesteps.
Therefore, we enable a seamless translation from the slice-wise model which processes slices individually, to a volume-wise model with better volumetric consistency.

\section{Experiments}

\para{Datasets.}
The experiments were conducted on two brain MRI datasets: SWI-to-MRA (S2M) dataset and RIRE~\cite{west1997comparison}\footnote{https://rire.insight-journal.org/index.html} T1-to-T2 dataset. 
To facilitate SWI-to-MRA brain MRI synthesis applications, we collected a high-quality SWI-to-MRA dataset. This dataset comprises paired SWI and MRA volume data of 111 patients that were acquired at Qilu Hospital of Shandong University using one 3.0T MRI scanner (\textit{i.e.,} Verio from Siemens). The SWI scans have a voxel spacing of $0.3438\times 0.3438\times 0.8$ mm and the MRA scans have a voxel spacing of $0.8984\times 0.8984\times 2.0$ mm. 
While most public brain MRI datasets lack high-quality details along z-axis and therefore are weak to indicate volumetric inconsistency, this volume data provides a good way to illustrate the performances for volumetric synthesis due to the clear blood vessels. We also evaluate our method on the public RIRE dataset~\cite{west1997comparison}. The RIRE dataset includes T1 and T2-weighted MRI volumes, and 17 volumes were used in the experiments. 

\para{Implementation Details.}
To summarize, for the S2M dataset, we randomly select 91 paired volumes for training and 20 paired volumes for inference; for the RIRE T1-to-T2 dataset, 14 volumes are randomly selected for training and 3 volumes are used for inference. 
All the volumes are resized to $256\times 256\times 100$ for S2M and $256\times 256\times 35$ for RIRE, where the last dimension represents the z-axis dimension, \textit{i.e.,} the number of slices in one volume for 2D image-to-image setting.
Our proposed method is built upon U-Net backbones. We use a pretrained KL autoencoder with a downsampling factor of $f = 4$. We train our model on an NVIDIA A100 80 GB GPU.

\para{Quantitative Results.}
We compare our pipeline to several baseline methods, including 2D-based methods: (1) Pix2pix~\cite{isola2017image}, a solid baseline for image-to-image translation; (2) Palette~\cite{saharia2022palette}, a diffusion-based method for 2D image translation; 3D-based methods: (3) a 3D version of Pix2pix, created by modifying the 2D backbone as a 3D backbone in the naive Pix2pix approach; and (4) a 3D version of CycleGAN~\cite{zhu2017unpaired}.
Naive 3D diffusion-based models are not included due to the lack of efficient backbones and the matter of timesteps' sampling efficiency.
We report the results in terms of mean absolute error (MAE), Structural Similarity Index (SSIM)~\cite{wang2004image}, and peak signal-to-noise ratio (PSNR).

Table~\ref{tbl:res} presents a quantitative comparison of our method and baseline approaches on the S2M and RIRE datasets. Our method achieves better performance than the baselines in terms of various evaluation metrics. To accelerate the sampling of diffusion models, we implement DDIM~\cite{song2020denoising} with 200 steps and report the results accordingly. It is worth noting that for the baseline approaches, the 3D version method (Pix2pix 3D) outperforms the corresponding 2D version (Pix2pix) at the cost of additional memory usage. For the Palette method, we implemented the 2D version but were unable to produce high-quality slices stably and failure cases dramatically affected the metrics results. Nonetheless, we included this method due to its great illustration of volumetric inconsistency.

\begin{table*}[t]
  \centering
  \caption{\textbf{Quantitative comparison on S2M and RIRE datasets.}}
  \begin{tabular}{lcccccccc}
    \toprule
     & \multicolumn{3}{c}{S2M} &  \multicolumn{3}{c}{RIRE~\cite{west1997comparison}} \\
    \cmidrule(r){2-4} \cmidrule(r){5-7}
    Methods & MAE $\downarrow$ & SSIM $\uparrow$ & PSNR $\uparrow$ & MAE $\downarrow$ & SSIM $\uparrow$ & PSNR $\uparrow$ \\
    \midrule
   Pix2pix~\cite{isola2017image}
      & 8.175  & 0.739 & 25.663 & 16.812 & 0.538 & 20.106\\
   Palette~\cite{saharia2022palette}  & 26.806 & 0.141 & 15.643 & 36.131 & 0.251 & 14.269\\
   Pix2pix 3D~\cite{isola2017image}  & 6.234  & 0.765 & 28.395 & 11.369 & 0.650 & 22.854\\
   CycleGAN 3D~\cite{zhu2017unpaired}   & 7.621  & 0.755 & 26.908 & 13.794 & 0.542 & 20.627\\
   \midrule
    \textbf{Ours 200 steps} & \textbf{5.243} & \textbf{0.788} & \textbf{29.446} & \textbf{10.794} & \textbf{0.676} & \textbf{24.332} \\
    \textbf{Ours 1000 steps} & \textbf{4.801}  & \textbf{0.801} & \textbf{30.143} & \textbf{10.619} & \textbf{0.684} & \textbf{25.458}  \\
    \bottomrule[1pt]
  \end{tabular}
  \label{tbl:res}
\end{table*}

\begin{figure}[ht]
\includegraphics[width=1.0\textwidth]{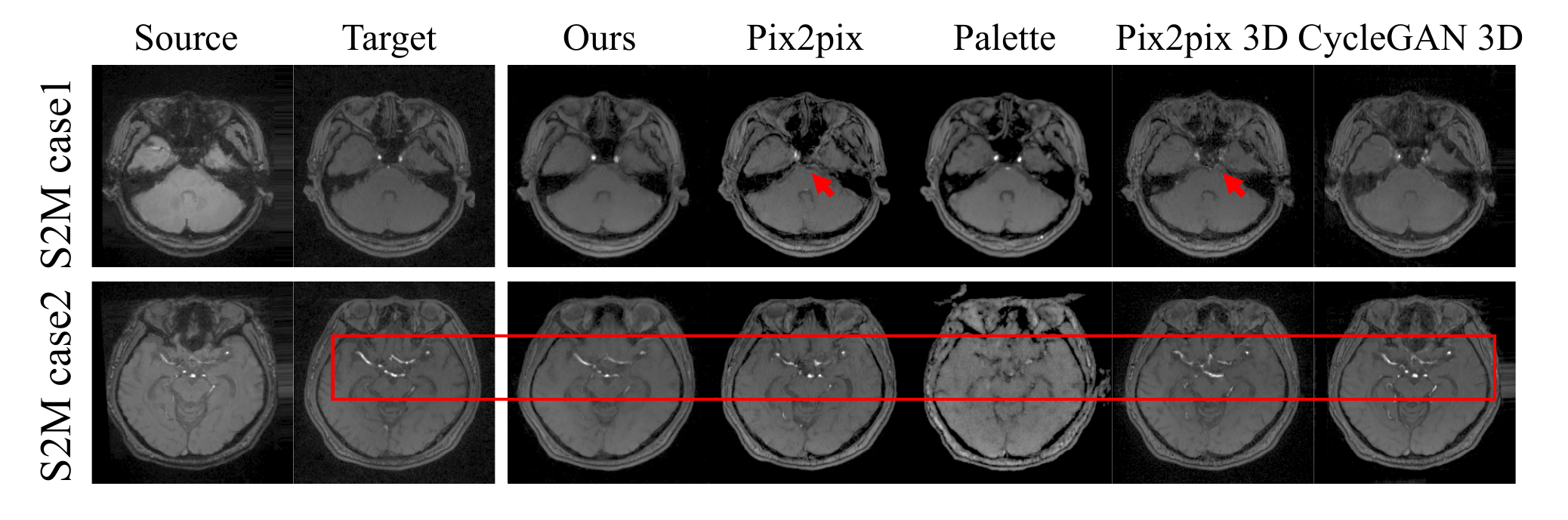}
\caption{\textbf{Qualitative comparison.} We compare our methods with baselines on two cases.}
\label{axial}
\end{figure}

\begin{figure}[ht]
\includegraphics[width=1.0\textwidth]{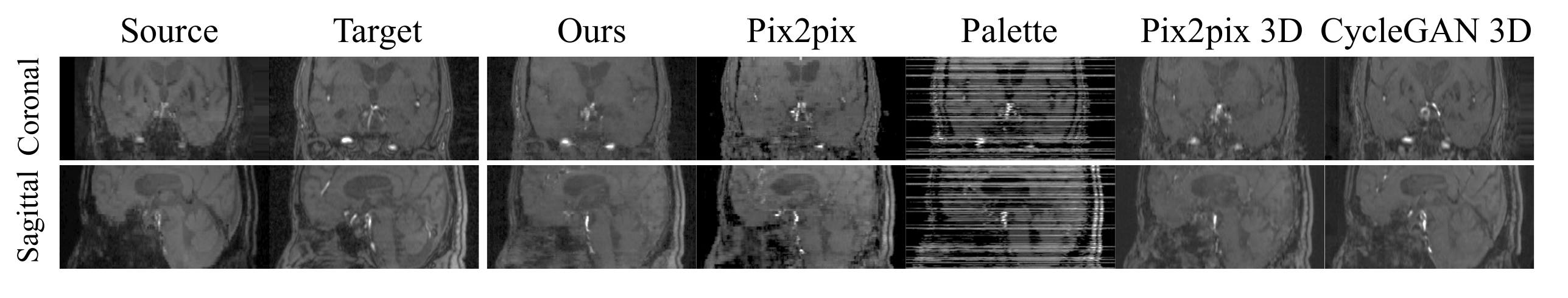}
\caption{\textbf{Coronal view and sagittal view.} To clearly indicate the volumetric consistency, we show a coronal view and a sagittal view of the volumes synthesized and the ground truth volumes.}
\label{volume_view}
\end{figure}

\para{Qualitative Results.}
Fig.~\ref{axial} presents a qualitative comparison of different methods, showcasing two axial slices of clear vessels. 
Our method synthesizes better images with more details, as shown in the qualitative results. The areas requiring special attention are highlighted with red arrows and red rectangles. It is worth noting that the synthesized axial slices not only depend on the source slice but also on the volume knowledge. 
For instance, for S2M case 1, the target slice shows a clear vessel cross-section that is based on the shape of the vessels in the volume.
In Fig.~\ref{volume_view}, we provide coronal and sagittal views. For methods that rely on 2D generation, we synthesize individual slices and concatenate them to create volumes. It is clear to observe the volumetric inconsistency examining the coronal and sagittal views of these volumes. For instance, Palette synthesizes 2D slices unstably, where some good slices are synthesized but others are of poor quality. 
As a result, volumetric inconsistency severely impacts the performance of volumes. While 2D baselines inherently introduce inconsistency in the coronal and sagittal views, 3D baselines also generate poor results than ours, particularly in regard to blood vessels and ventricles.

\begin{table*}[ht]
  \centering
  \caption{\textbf{Ablation Quantitative Results.}}
  \begin{tabular}{lcccccccc}
    \toprule
     & \multicolumn{3}{c}{S2M} &  \multicolumn{3}{c}{RIRE~\cite{west1997comparison}} \\
    \cmidrule(r){2-4} \cmidrule(r){5-7}
    Methods & MAE $\downarrow$ & SSIM $\uparrow$ & PSNR $\uparrow$ & MAE $\downarrow$ & SSIM $\uparrow$ & PSNR $\uparrow$ \\
    \midrule
    \textbf{w/o volumetric layers} & 5.128 & 0.792 & 29.894 & 10.925 & 0.667 & 24.623 \\
    \textbf{w/ volumetric layers} & 4.801 & 0.801 & 30.143 & 10.619 & 0.684 & 25.458  \\
    \bottomrule[1pt]
  \end{tabular}

  \label{tbl:abl}
\end{table*}

\begin{figure}[t]
\centering
\includegraphics[width=0.7\textwidth, height=0.35\textwidth]{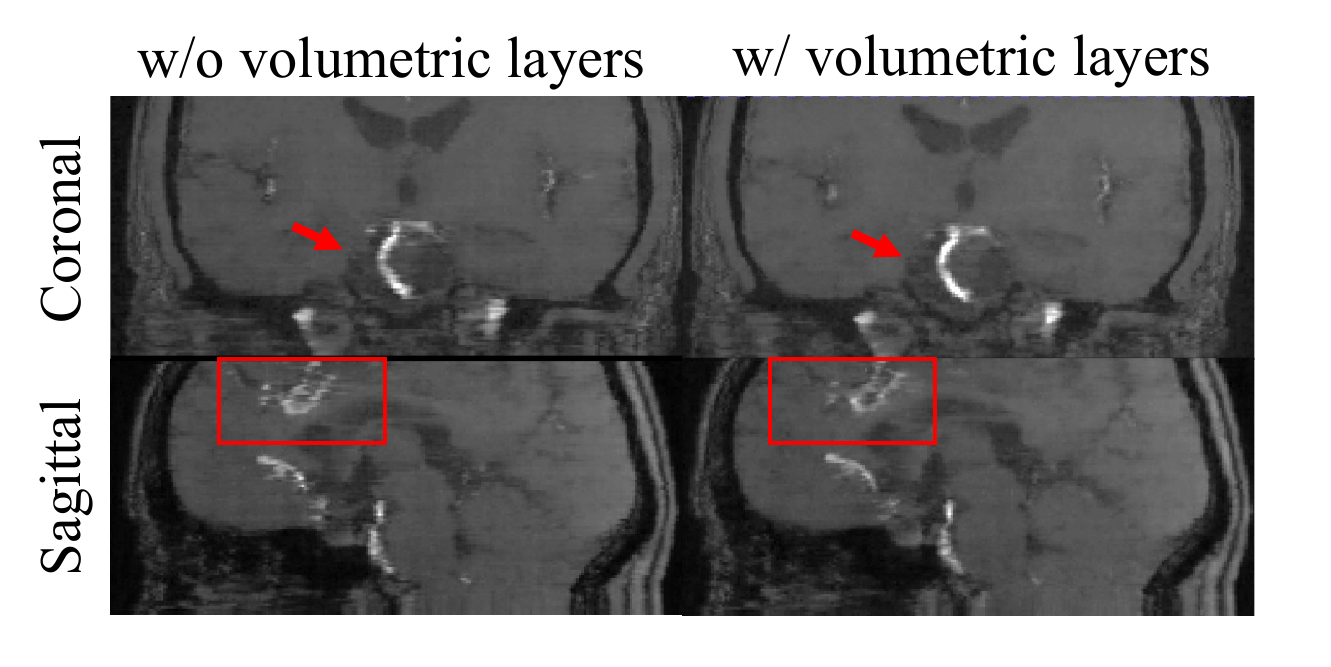}
\caption{\textbf{Ablation qualitative results with coronal view and sagittal view.}}
\label{figure_abl}
\end{figure}

\para{Ablation Analysis.} 
We conduct an ablation study to show the effectiveness of volumetric fine-tuning. Table~\ref{tbl:abl} presents the quantitative results, demonstrating that our approach is able to increase the model's performance beyond that of the slice-wise model, without incurring significant extra training expenses. 
Fig.~\ref{figure_abl} illustrates that fine-tuning volumetric layers helps to mitigate volumetric artifacts and produce clearer vessels, which is crucial for medical image synthesis.

\section{Conclusion}
In this paper, we propose Make-A-Volume, a diffusion-based framework for cross-modality 3D medical image synthesis. Leveraging latent diffusion models, our method achieves high performance and can serve as a strong baseline for multiple cross-modality medical image synthesis tasks. 
More importantly, we introduce a generic paradigm for volumetric data synthesis by utilizing 2D backbones and demonstrate that fine-tuning volumetric layers helps the two-stage model capture 3D information and synthesize better images with volumetric consistency. 
We collected an in-house SWI-to-MRA dataset with clear blood vessels to evaluate volumetric data quality.
Experimental results on two brain MRI datasets demonstrate that our model achieves superior performance over existing baselines. 
Generating coherent 3D and 4D data is at an early stage in the diffusion models literature, we believe that by leveraging slice-wise models and extending them to 3D/4D models, more work can help achieve better volume synthesis with reasonable memory requirements. 
In the future, we will investigate more efficient approaches for more high-resolution volumetric data synthesis.

\section*{Acknowledgement} The work described in this paper was partially supported by grants from the Research Grants Council of the Hong Kong Special Administrative Region, China (T45-401/22-N), the National Natural Science Fund (62201483), HKU Seed Fund for Basic Research (202009185079 and 202111159073), RIE2020 Industry Alignment Fund – Industry Collaboration Projects (IAF-ICP) Funding Initiative, as well as cash and in-kind contribution from the industry partner(s).

\bibliographystyle{splncs04}
\bibliography{reference}

\end{document}